\documentclass[fleqn,12pt,twoside]{article}
\usepackage{espcrc1}

\usepackage{graphicx}

\newcommand{\AmS}{{\protect\the\textfont2
  A\kern-.1667em\lower.5ex\hbox{M}\kern-.125emS}}

\title{The Theory of Nuclear Forces:\\
Is the Never-Ending Story Coming to an End?}

\author{R. Machleidt\address{Department of Physics, University of Idaho, Moscow, Idaho
83844, USA}
\thanks{Plenary talk presented at the 18th International Conference
on Few-Body Problems in Physics, August 21-25, 2006, Santos, SP, Brazil;
to be published in Nucl.\ Phys.\ A.}}

\begin{document}

\maketitle

\begin{abstract}
I review the recent progress in our
understanding of nuclear forces in terms of an effective
field theory (EFT) for low-energy QCD
and put this progress into historical perspective.
This is followed by an assessment of the current status
of EFT based nuclear potentials.
In concluding, I will summarize some unresolved issues.
\end{abstract}

\section {Historical Perspective}

The theory of nuclear forces has a long history.
Based upon the Yukawa idea~\cite{Yuk35} and the discovery of the pion,
the 1950's became the first period of ``pion theories''.
These, however, resulted in failure---for reasons we
understand today: pion dynamics is ruled by
chiral symmetry, a constraint that was not realized
in the theories of the 1950's.

The 1960's and 70's represent the main period for
theories that also include heavy mesons
(``meson theories'')~\cite{Mac89}, but the work on meson models continued 
all the way into the 1990's when the family of
the so-called high-precision NN potentials was developed.
This family includes
the Nijm-I, Nijm-II, and Reid93 potentials~\cite{Sto94}, 
the Argonne $V_{18}$~\cite{WSS95}, and
the CD-Bonn potential~\cite{MSS96,Mac01}. 
Later, also the highly non-local potential by
Doleschall {\it et al.}~\cite{Dol03} and the
``CD-Bonn + $\Delta$'' model by 
Deltuva {\it et al.}~\cite{DMS03}
joined the club.
All these potentials have in common that they are charge-dependent,
use about 40-50 parameters, and reproduce
the 1992 Nijmegen NN data
base with a $\chi^2$/datum $\approx 1$.

Over the past ten years, the high-precision
potentials have been applied intensively
in exact few-body calculations and miscroscopic
nuclear structure theory.
Already the first few applications of the high-precision
potentials in three-nucleon reactions~\cite{Glo96} clearly
revealed sizable differences 
between the predictions from 
NN potentials with a $\chi^2$/datum $\approx 1$ (i.e., the new
generation of potentials) 
and NN potentials with a $\chi^2$/datum $\approx 2$  
(the old generation of the 1970's and 80's which includes
the old Nijmegen, the Paris, 
and the old Bonn potentials). 
Thus, once for all,
the standard of precision was established that must be met
by any future work in microscopic nuclear structure and exact
few-body calculations: The input NN potential must
reproduce the NN data with a
 $\chi^2$/datum $\approx 1$ or the uncertainty in the predictions
will make it impossible to draw
reliable conclusions.

In spite of these great practical achievements, the
high-precision potentials cannot be the end of the story,
because they are all phenomenological in nature.
Ultimately, we need potentials that are based on
proper theory {\it and\/} yield quantitative results.

Since the nuclear force is a manifestation of strong interactions,
any serious derivation has to start from
quantum chromodynamics (QCD).
However, the well-known problem with QCD is 
that it is
non-perturbative in the low-energy
regime characteristic for nuclear physics.
For many years, this fact was perceived as a major obstacle for
a derivation of the nuclear force from QCD---impossible to overcome
except by lattice QCD.
The effective field theory (EFT) concept has shown the
way out of this dilemma. 
One has to realize that the scenario of low-energy QCD 
is characterized by pions
and nucleons interacting via a force governed by spontaneously broken
approximate chiral symmetry. 

Based upon this EFT, a systematic expansion
in terms of $(Q/\Lambda_\chi)^\nu$ can be developed,
where $Q$ denotes a momentum or pion mass, 
$\Lambda_\chi \approx 1$ GeV is the chiral symmetry breaking
scale, and $\nu \geq 0$~\cite{Wei90}.
This has become known as chiral perturbation theory (ChPT).
In contrast to the pion theories of the 1950's,
this new chirally constrained ``pion theory'' is quite successful.
Starting with the pioneering work by
Ord\'o\~nez, Ray, and van Kolck
in the early 1990's~\cite{ORK94}, NN potentials have been constructed
in the framework of ChPT. 
In the next section,
I will discuss the latest
status of this work. 

A remarkable fact of ChPT is that it makes specific
predictions also for many-body forces. For a given order of ChPT,
two-nucleon forces (2NF), three-nucleon forces
(3NF), \ldots are generated on the same footing
(see Ref.~\cite{ME05} for a concise review).
At leading order (LO), there are no 3NF, and
at next-to-leading order (NLO),
all 3NF terms cancel~\cite{Wei90,Kol94}. 
However, at next-to-next-to-leading order (NNLO) and higher orders, well-defined, 
nonvanishing 3NF occur~\cite{Kol94,Epe02b}.
Since 3NF show up for the first time at NNLO, they are weak.
Four-nucleon forces (4NF) occur first at N$^3$LO and are, therefore,
even weaker.

\begin{table}[t]
\caption{\small $\chi^2$/datum for the reproduction of the 1999 $np$ database
by families of $np$ potentials at NLO and NNLO constructed by the
Bochum/Juelich group~\cite{EGM04}.}
\small
\begin{tabular}{cccccc}
\hline 
\hline 
 &  \# of $np$ && \multicolumn{3}{c}{\it Bochum/Juelich}\\
 Bin (MeV) 
 & data 
 &
 & NLO
 & 
 & NNLO 
\\
\hline 
\hline 
0--100&1058&&4--5&&1.4--1.9\\ 
100--190&501&&77--121&&12--32\\ 
190--290&843&&140--220&&25--69\\ 
\hline 
0--290&2402&&67--105&&12--27
\\ 
\hline 
\hline 
\end{tabular}
\end{table}

\section{Current status of chiral NN potentials}

As pointed out in the previous section,
the research conducted during the high-precision period of the 1990's~\cite{Glo96}
systematically tested and firmly established the standards
of precision that must be met by NN potentials to
yield sufficiently accurate predictions when applied in microscopic
calculations of few-body systems.
Since in nuclear EFT we are dealing with a perturbative expansion (ChPT),
the question is to what order of ChPT we have to go
to obtain the precision that meets those standards.
To discuss this issue on firm grounds, I show in Table~1
the $\chi^2$/datum for the fit of the world $np$ data
below 290 MeV for a family of $np$ potentials at 
NLO and NNLO. 
The NLO potentials produce the horrendous $\chi^2$/datum between 67 and 105,
and the NNLO are between 12 and 27.
The rate of improvement from one order to the other
is very encouraging, but the quality of the reproduction
of the $np$ data at NLO and NNLO is obviously totally
insufficient for reliable predictions.
In the literature, one can find calculations employing
NLO and NNLO potentials in few-nucleon systems. Because of the extremely poor
accuracy of such potentials, no reliable conclusions can be
drawn from such calculations, rendering
these applications essentially useless.

\begin{table}[t]
\caption{$\chi^2$/datum for the reproduction of the 1999 {\boldmath\bf $np$ database}
by various $np$ potentials.
Numbers in parentheses denote cutoff parameters in units of MeV.}
\small
\begin{tabular}{cc|c|c|c}
\hline 
\hline 
 &  
 & {\it Idaho}
 & {\it Bochum/Juelich}
 & Argonne         
\\
 Bin (MeV)
 & \# of {\boldmath $np$}
 & N$^3$LO~\cite{EM03}
 & N$^3$LO~\cite{EGM05} 
 & $V_{18}$~\cite{WSS95}
\\
 &  data
 & (500--600)
 & (600/700--450/500)
 & 
\\
\hline 
\hline 
0--100&1058&1.0--1.1&1.0--1.1&0.95\\ 
100--190&501&1.1--1.2&1.3--1.8&1.10\\ 
190--290&843&1.2--1.4&2.8--20.0&1.11\\ 
\hline 
0--290&2402&1.1--1.3&1.7--7.9&1.04
\\ 
\hline 
\hline 
\end{tabular}
\end{table}

\begin{table}[b]
\caption{$\chi^2$/datum for the reproduction of the 1999 {\boldmath\bf $pp$ database}
by various $pp$ potentials.
Numbers in parentheses denote cutoff parameters in units of MeV.}
\small
\begin{tabular}{cc|c|c|c}
\hline 
\hline 
              
 &  
 & {\it Idaho}
 & {\it Bochum/Juelich}
 & Argonne         
\\
 Bin (MeV)
 & \# of {\boldmath $pp$}
 & N$^3$LO~\cite{EM03}
 & N$^3$LO~\cite{EGM05} 
 & $V_{18}$~\cite{WSS95}
\\
 &  data
 & (500--600)
 & (600/700--450/500)
 & 
\\
\hline 
\hline 
0--100&795&1.0--1.7&1.0--3.8&1.0 \\ 
100--190&411&1.5--1.9&3.5--11.6&1.3 \\ 
190--290&851&1.9--2.7&4.3--44.4&1.8 \\ 
\hline 
0--290&2057&1.5--2.1&2.9--22.3&1.4 
\\ 
\hline 
\hline 
\end{tabular}
\end{table}

Based upon these facts, it has been pointed out in 2002 by
Entem and Machleidt~\cite{EM02a,EM02} that NNLO is insufficient and one has
to proceed to N$^3$LO. Consequently, the first N$^3$LO  potential was
created in 2003~\cite{EM03}, which showed that at this order
a $\chi^2$/datum comparable to the high-precision
Argonne $V_{18}$ potential can, indeed, be achieved, see Table~2.
This ``Idaho'' N$^3$LO potential~\cite{EM03} produces
a $\chi^2$/datum = 1.1 
for the world $np$ data below 290 MeV
which compares well with the $\chi^2$/datum = 1.04
by the Argonne potential.
In 2005, also the Bochum/Juelich group produced
several N$^3$LO NN potentials~\cite{EGM05}, the best of which
fits the $np$ data with
a $\chi^2$/datum = 1.7 and the worse with 
a $\chi^2$/datum = 7.9 (see Table~2).
While 7.9 is clearly unacceptable for any meaningful
application, a $\chi^2$/datum of 1.7 is reasonable,
although it does not meet
the precision standard established in the 1990's.

I turn now to the $pp$ data.
Typically, $\chi^2$ for $pp$ data are larger than for $np$
because of the higher precision of $pp$ data.
Thus, the Argonne $V_{18}$ produces
a $\chi^2$/datum = 1.4 for the world $pp$ data
below 290 MeV and the best Idaho N$^3$LO $pp$ potential obtains
1.5. The fit by the best Bochum/Juelich 
N$^3$LO $pp$ potential results in
a $\chi^2$/datum = 2.9 which, again, 
is not quite consistent with the precision standards of the 1990's.
The worst Bochum/Juelich N$^3$LO $pp$ potential produces
a $\chi^2$/datum of 22.3 and 
is incompatible with realiable predictions.

Phase shifts of $np$ scattering from the best Idaho 
(solid line) and Bochum/Juelich (dashed line)
N$^3$LO $np$ potentials are shown in Figure~1.
The phase shifts confirm what the corresponding
$\chi^2$'s have already revealed.

\begin{figure}[t]
\vspace*{-1.0cm}
\hspace*{-1.5cm}
\scalebox{0.45}{\includegraphics{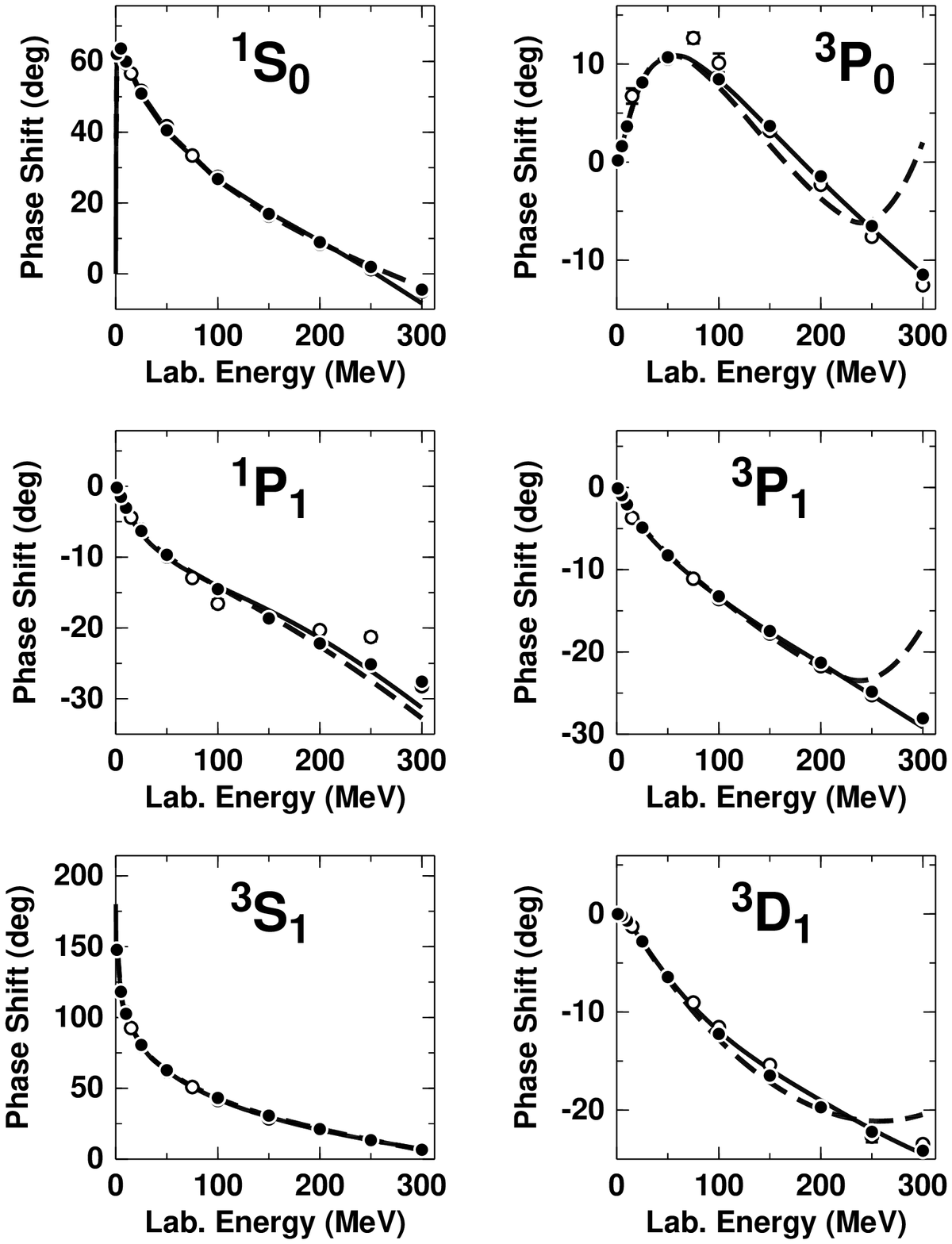}}
\hspace*{-2.0cm}
\scalebox{0.45}{\includegraphics{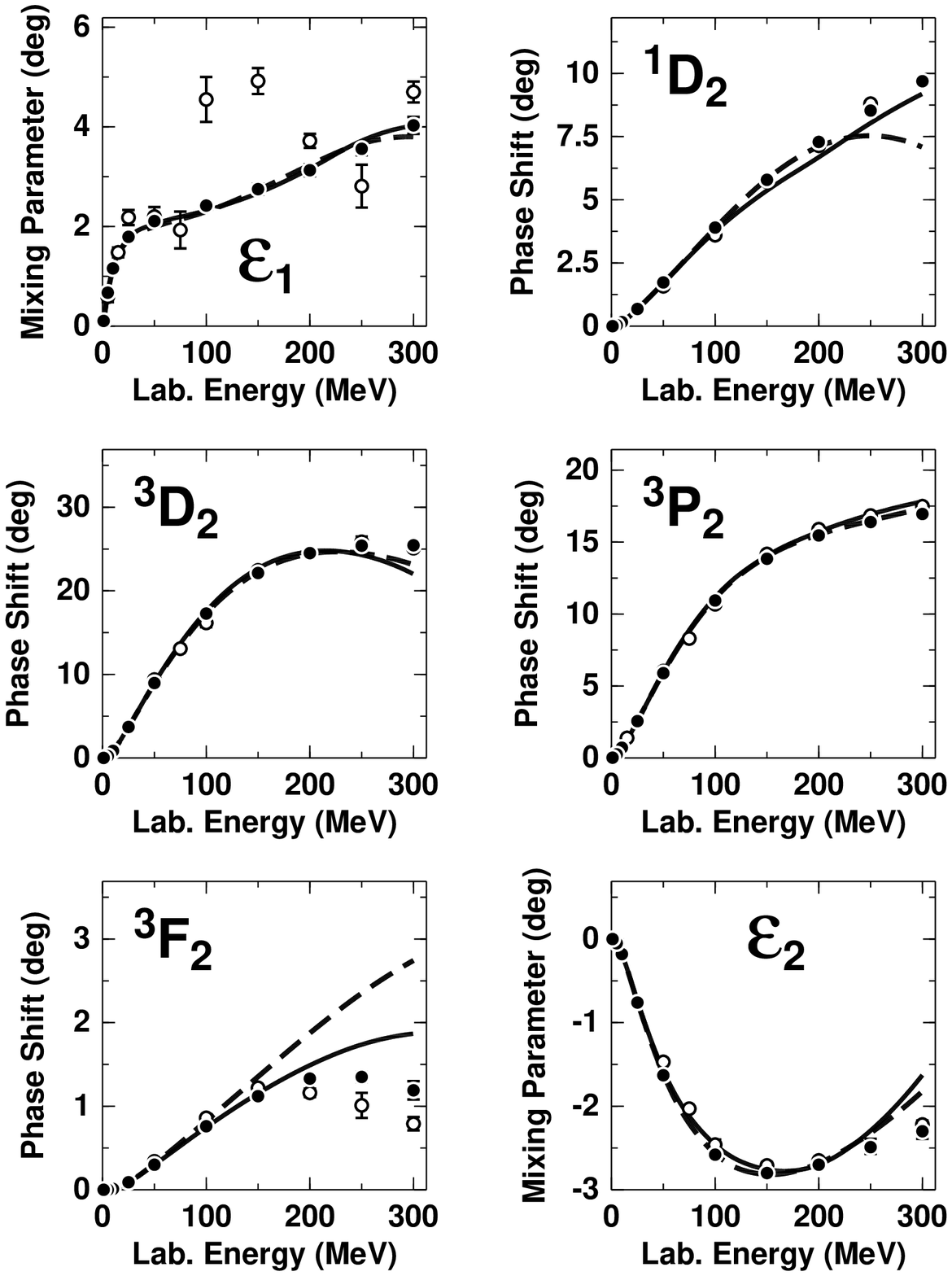}}
\vspace*{-3.0cm}
\caption{\small $np$ phase shifts for energies below 300 MeV and partial
waves with total angular momentum $J\leq 2$.
The solid and the dashed curves show the phase shifts
predicted by the N$^3$LO potentials constructed by the
Idaho~\cite{EM03} and the Bochum/Juelich~\cite{EGM05}
groups, respectively. Solid dots and open circles represent the Nijmegen
multienergy $np$ phase shift analysis~\cite{Sto93} and the
GWU/VPI single-energy $np$ analysis SM99~\cite{SM99}, respectively.}
\end{figure}

\section{The low-energy constants}

\begin{table}
\caption{Low-energy parameters.
$c_i$ are given in units of GeV$^{-1}$ 
and $\bar{d}_i$ in units GeV$^{-2}$.}
\small
\begin{tabular}{crrrc}
\hline
\hline
 &             & {\it Idaho}~\cite{EM03} & {\it Boch./J.}\cite{EGM05} & Nijmegen \\
 & $\pi N$~\cite{BM00,FMS98} & N$^3$LO pot. & N$^3$LO pot. & 2003 PWA~\cite{RTS03} \\
 &                      &              &                  & (1999 PWA~\cite{Ren99})   \\
\hline
\hline
$c_1$ & $-0.81\pm 0.15$ & --0.81 & --0.81 & --0.76 \\ 
                                       &&&& (--0.76)\\
$c_2$ & $3.28\pm 0.23$  & 2.80 & 3.28 & -- \\ 
      & & & & (--) \\
$c_3$ & $-4.69\pm 1.34$ & --3.20 & --3.40 & $-4.78\pm 0.10$\\ 
                                    & & & & $(-5.08\pm 0.28)$\\
$c_4$ & $3.40\pm 0.04$  &   5.40 &   3.40 & $3.96\pm 0.22$\\ 
                                    & & & & $(4.70\pm 0.70)$\\
\hline 
$\bar{d}_1 + \bar{d}_2$  & $3.06\pm 0.21$ & 3.06 & 3.06 & -- \\ 
$\bar{d}_3$ & $-3.27\pm 0.73$ & --3.27 & --3.27 & -- \\ 
$\bar{d}_5$ & $0.45\pm 0.42$ & 0.45 & 0.45 & -- \\ 
$\bar{d}_{14} - \bar{d}_{15}$ & $-5.65\pm 0.41$ & --5.65 & --5.65 & -- \\ 
\hline
\hline
\end{tabular}
\end{table}

\begin{figure}[t]
\vspace*{-0.5cm}
\scalebox{0.25}{\includegraphics{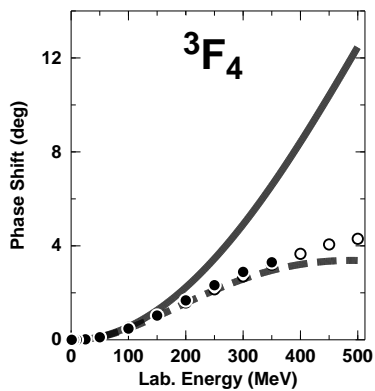}}

\vspace*{-6.5cm}
\hspace*{6.0cm}
\begin{minipage}{6cm} 
\caption{\small $^3F_4$ phase shifts calculated at NNLO using
$c_3 = -4.78$ GeV$^{-1}$. The solid curve is based upon the
NNLO amplitude whithout cutoff, while for the dashed curve the cutoff
used in the Nijmegen analysis~\cite{RTS03} is employed.
Solid dots and open circles like in Fig.~1.}
\end{minipage}

\end{figure}

The two-pion-exchange contribution to the NN interaction
depends on the low-energy constants (LECs) of the dimension-two 
Lagrangian, commonly denoted by $c_i$, and the dimension-three
LECs, $d_i$~\cite{EM02}. 
These parameters have been determined
in $\pi N$ data analyses, see column $\pi N$ of
Table 4. 
The most influential LEC is $c_3$ and, so, we will focus
here just on $c_3$.
Not surprisingly, the Idaho and Bochum/Juelich N$^3$LO
potentials apply essentially the same value, namely,
$c_3 = -3.3\pm 0.1$ 
GeV$^{-1}$, 
which is on the lower side
but still roughly within one standard deviation of its $\pi N$
determination. 
There exists also a determination
of this parameter from the NN data by the Nijmegen group.
They report 
$c_3 = -5.08\pm 0.28$ 
GeV$^{-1}$ 
in their 1999 $pp$ analysis~\cite{Ren99} and 
$c_3 = -4.78\pm 0.10$ 
GeV$^{-1}$ 
in their 2003 $pp+np$ analysis~\cite{RTS03}.
Knowing the sensitivity of NN phase shifts and
observables to $c_3$, the difference
between the NN potential values of about --3.3
GeV$^{-1}$ 
and the Nijmegen NN analysis value of --4.8 
GeV$^{-1}$ 
is {\it huge}.
Off-hand, it is hard to understand how two different types of NN analyses 
can be in such severe disagreement.

However, the discrepancy is easily explained if one looks into the
details of the Nijmegen analysis~\cite{EM03a}.
The Nijmegen group starts from the NN amplitude at NNLO in momentum
space, Fourier transforms it into $r$-space to obtain
a local potential, and then cuts this potential off
at r=1.6 fm (i.e., the potential is set to zero 
for $r\leq 1.6$ fm). The impact of this cutoff
is demonstrated in Figure~2 for
the $^3F_4$ phase shifts, which is a representative
case. In that figure, the solid line shows
the phase shifts derived from the NN amplitude at NNLO
using 
$c_3 = -4.78$ 
GeV$^{-1}$ 
and applying no cutoffs (i.e., 
the actual model-independent amplitude at NNLO is used).
The dashed line is obtained when the Nijmegen cutoff
is applied and, as before, 
$c_3 = -4.78$
GeV$^{-1}$ is used. 
The difference between the two curves in Fig.~2 is entirely
due to the cutoff applied in the Nijmegen analysis.
It is clearly seen that the influence of this cutoff is
dramatic and introduces a huge systematic error
rendering the Nimegen analysis and their results
unreliable~\cite{EM03a}.

It must be noted that the Idaho and Bochum/Juelich N$^3$LO potentials
also use cutoffs, but not as harsh ones as in
the Nijmegen analysis.

The best way to determine the LECs from NN data is 
to apply no cutoffs to the NN amplitudes (at NNLO
or N$^3$LO) in
peripheral partial waves and to keep the amplitudes of
lower partial waves at their empirical values (as determined in
phase shift analysis).
In such an analysis the value
$c_3 = -2.7$ 
GeV$^{-1}$ 
is obtained~\cite{EM02b}.

\section{Conclusions}

The theory of nuclear forces has made great progress
since the turn of the millennium.
Nucleon-nucleon potentials have been developed that are based 
on proper theory (EFT for low-energy QCD) 
and are of high-precision, at the same time. 
Moreover, the theory generates
two- and many-body forces on an equal footing
and provides a theoretical explanation for
the empirically known fact that 2NF $\gg$ 3NF $\gg$ 4NF
\ldots.

At N$^3$LO~\cite{EM02,EM03}, the accuracy can be achieved that
is necessary and sufficient for microscopic nuclear structure.
First calculations applying the N$^3$LO
NN potential~\cite{EM03} in the (no-core) shell model 
\cite{Cor02,Cor05,NC04,FNO05,Var05},
the coupled cluster formalism
\cite{Kow04,DH04,Wlo05,Dea05,Gou06},
and the unitary-model-operator approach~\cite{FOS04}
have produced promising results. 

The 3NF at NNLO is known~\cite{Kol94,Epe02b} and has been
applied in few-nucleon reactions~\cite{Epe02b,Erm05,Wit06}
as well as the structure of light nuclei~\cite{Nog04,Nog06}. 
However, the famous `$A_y$ puzzle' of nucleon-deuteron
scattering is not resolved by the 3NF at NNLO. Thus, the 
most important outstanding issue is the 3NF at N$^3$LO,
which is under construction~\cite{Mei06}.

Another open question that needs to be settled 
is whether Weinberg power counting, which is applied in
all current NN potentials, is consistent. This controversial issue 
is presently being debated in the literature~\cite{NTK05,EM06,Kol06}.

It may be too early to claim that the never-ending-story
is coming to an end, but the story seems to be converging---at the
same rate as chiral perturbation theory.

\section*{Acknowledgements}
This work was supported by the U.S. National Science
Foundation under Grant No.~PHY-0099444.

\end{document}